%% file: openvocab_nlm_icse2019.tex
\documentclass[sigconf]{acmart}

\copyrightyear{2020} 
\acmYear{2020} 
\setcopyright{rightsretained} 
\acmConference[ICSE '20]{42nd International Conference on Software Engineering}{May 23--29, 2020}{Seoul, Republic of Korea}
\acmBooktitle{42nd International Conference on Software Engineering (ICSE '20), May 23--29, 2020, Seoul, Republic of Korea}
\acmDOI{10.1145/3377811.3380342}
\acmISBN{978-1-4503-7121-6/20/05}

\PassOptionsToPackage{hyphens}{url}
\usepackage{hyperref}
\usepackage{amsmath,amssymb,amsfonts}
\usepackage{algorithm}
\usepackage[noend]{algpseudocode}
\usepackage{graphicx}
\usepackage{textcomp}
\usepackage{xcolor}
\usepackage{xspace}
\usepackage{url}
\usepackage{tikz}
\usepackage{filecontents}
\usepackage{siunitx}
\DeclareSIUnit{\million}{M}
\DeclareSIUnit{\billion}{B}
\DeclareSIUnit{\thousand}{K}
\usepackage{paralist}

\usepackage{multirow}

\usepackage{booktabs}
\usepackage[normalem]{ulem}
\usepackage{tcolorbox}
\usepackage{enumerate}
\usepackage[shortlabels]{enumitem}

\definecolor{pblue}{rgb}{0.13,0.13,1}
\definecolor{pgreen}{rgb}{0,0.5,0}
\definecolor{pred}{rgb}{0.9,0,0}
\definecolor{pgrey}{rgb}{0.46,0.45,0.48}
\definecolor{java}{rgb}{0.2, 0.4, 0.9}
\usepackage{listings}
\input{dois.tex}
\begin{document}

\newcommand{\eot}{\texttt{</t>}\xspace}
\newcommand{\id}[1]{\texttt{#1}}
\algrenewcommand\algorithmicindent{0.75em}%

\newcommand*\conclusion[1]{
\begin{tcolorbox}[left=0mm,right=0mm,boxrule=0.25mm,colback=gray!5!white]
\vspace{-0.2cm}
#1
\vspace{-0.2cm}
\end{tcolorbox}
}

\renewcommand{\paragraph}[1]{\emph{#1}}


\title[Big Code != Big Vocabulary]{Big Code != Big Vocabulary: \\ Open-Vocabulary Models
for Source Code}

\author{Rafael-Michael Karampatsis}
 \affiliation{
   \institution{University of Edinburgh}
   \city{Edinburgh}
   \country{United Kingdom}
}
\author{Hlib Babii}
\affiliation{
    \institution{Free University of Bozen-Bolzano}
    \city{Bozen-Bolzano}
    \country{Italy}
}

\author{Romain Robbes}
\affiliation{
\institution{Free University of Bozen-Bolzano}
 \city{Bozen-Bolzano}
    \country{Italy}
}

\author{Charles Sutton}
 \affiliation{
   \institution{Google AI, University of Edinburgh and The Alan Turing Institute}
   \city{Mountain View, CA}
   \country{United States}
}

\author{Andrea Janes}
\affiliation{
\institution{Free University of Bozen-Bolzano}
 \city{Bozen-Bolzano}
    \country{Italy}
}

\newboolean{showcomments}
\ifthenelse{\boolean{showcomments}}{%
   \newcommand{\maybecomment}[1]{ {\lbrack #1 \rbrack} }
   }{
   \newcommand{\maybecomment}[1]{ }
}

\renewcommand{\shortauthors}{R. Karampatsis et al.}

\newcommand{\todo}[1]{{\color{red}\maybecomment{TODO: #1}}}
\newcommand{\cs}[1]{{\color{purple}\maybecomment{From Charles: #1}}}
\newcommand{\raf}[1]{{\color{blue}\maybecomment{From Rafa: #1}}}
\newcommand{\hlib}[1]{{\color{yellow}\maybecomment{From Hlib: #1}}}
\newcommand{\rr}[1]{\textbf{\maybecomment{RR: #1}}}
\newcommand{\etal}[0]{\textit{et al.}\xspace}
\newcommand{\ra}[1]{\renewcommand{\arraystretch}{#1}}

\newcommand*\choices[0]{\textbf{Choices. }}
\newcommand*\nonEnglish[0]{\texttt{<non-en>}\xspace}
\newcommand*\unk[0]{\texttt{<unk>}\xspace}
\newcommand*\underscore[0]{\texttt{<\_>}\xspace}
\newcommand*\Uppercase[0]{\texttt{<Upper>}\xspace}
\newcommand*\UPPERCASE[0]{\texttt{<UPPER>}\xspace}
\newcommand*\tokstring[0]{\texttt{<string>}\xspace}
\newcommand*\tokcomment[0]{\texttt{<comment>}\xspace}
\newcommand*\nostring[0] {\sout{\texttt{"str"}}}
\newcommand*\nocomment[0] {\sout{\texttt{/**/}}}
\newcommand*\yesstring[0] {\texttt{"str"}}
\newcommand*\yescomment[0] {\texttt{/**/}}
\newcommand*\yeswhitespace[0] {\texttt{<tab>}}
\newcommand{\spl}[0]{\begin{picture}(2,0)(-1,-3)\circle*{3}\end{picture}\ }

\newcommand*\nowhitespace[0] { \yescomment \ \  \yesstring}

\newcommand*{\numproj}{14,436\xspace}
\newcommand*{\numprojtrain}{13,362\xspace}
\newcommand*{\numprojtest}{2,165\xspace}
\newcommand*{\numprojval}{2,165\xspace}

\newcommand* \splitnum[0] {\texttt{1/2/3}}

\begin{abstract}
Statistical language modeling techniques have successfully been applied to large source code corpora, yielding a variety of new software development tools, such as tools for code suggestion, improving readability, and API migration.
A major issue with these techniques is that code introduces new vocabulary at a far higher rate than natural language, as new identifier names proliferate. Both large vocabularies and out-of-vocabulary issues severely affect Neural Language Models (NLMs) of source code, degrading their performance and rendering them unable to scale. \\
In this paper, we address this issue by: 1) studying how various modelling choices impact the resulting vocabulary on a large-scale corpus of \numprojtrain projects; 2) presenting an \emph{open vocabulary} source code NLM that can scale to such a corpus, 100 times larger than in previous work; and 3) showing that such models outperform the state of the art on three distinct code corpora (Java, C, Python). To our knowledge, these are the largest NLMs for code that have been reported. \\
All datasets, code, and trained models used in this work are publicly available. 

\end{abstract}



\maketitle

\section{Introduction}

Many works have taken advantage of the ``naturalness'' of software \cite{Hindle2012} to assist software engineering tasks, including code completion \cite{Raychev2014}, improving
code readability \cite{Allamanis2014}, program repair \cite{santos2018syntax,chen2018sequencer}, identifying buggy code \cite{Ray2016} 
and API migration \cite{gu2017deepam}, among many others \cite{big-code-survey}. These approaches analyze large amounts of source code, ranging from hundreds to thousands of software projects, building machine learning models of source code properties, inspired by techniques from natural language processing (NLP).

When applying any NLP method to create any type of
software development tool, a crucial early decision is how to model software's vocabulary. This is all the more important because, unlike in natural language, \textbf{software developers are free to create any identifiers they like, and can make them arbitrarily complex}. Because of this
fundamental fact, any model that is trained on a large-scale software corpus has to deal with an extremely large and sparse vocabulary (Section \ref{sec:background}). Rare words can not be modelled effectively. Furthermore, if identifiers were not observed in the training set, 
many classes of models cannot predict them, which is known as the \emph{out-of-vocabulary (OOV) problem}. Hellendoorn and Devanbu observe this issue 
for the task of language modeling, showing that a neural language model (NLM) has difficulties scaling beyond as few as a hundred projects \cite{Hellendoorn2017}.
Given that neural approaches are the state-of-the-art in NLP,
finding ways to scale them to a larger software corpus is a very important goal. 

Our \emph{first contribution} is a thorough study of
the effects of the vocabulary design choices
that must be made when creating any NLP model of software (Section \ref{sec:choices}).
The vocabulary design choices we study include how
to handle comments, string literals, and white space; 
whether to filter out infrequent tokens; and whether and how
to split \emph{compound tokens}, such as names that contain camel
case and underscores.
We examine how these choices affect
the vocabulary size, which affects the scalability
of models, and how they affect the OOV rate, that is, how often
the vocabulary fails to include names that appear
in new projects.
We find that the choices have a large impact, leading to variations in vocabulary size of up to \emph{three orders of magnitude}. However, we find that the most common ways to reduce vocabulary that were previously considered in the software engineering literature, such as splitting identifiers according to underscores and case, are not enough to obtain a vocabulary of a manageable size; advanced approaches such as adaptations of the Byte-Pair Encoding (BPE) algorithm \cite{Gage1994,Sennrich2015} are needed to reach this goal and deal with the OOV problem.

This empirical study motivates our \emph{second contribution}. Drawing on our results, we develop a large-scale open-vocabulary NLM for source code (Section \ref{sec:rnnlm}).
To our knowledge, this is the first BPE NLM for source code reported in the literature.
This NLM model leverages BPE, beam search, and caching to both keep vocabulary size low and successfully predict OOV tokens. We show that this NLM is able to scale: we train it on up to \numprojtrain software projects, yielding the largest NLM trained on source code we are aware of. 

Finally, in our \emph{third contribution} we extensively evaluate our NLM (Sections \ref{sec:evaluation}--\ref{sec:results}). We show that the open-vocabulary NLM
outperforms both
$n$-gram LMs and closed vocabulary NLMs for the task of code completion for several languages (Java, C, and Python).
To show that  improvement in language modelling transfers
to downstream SE tasks, we conduct an experiment similar to 
\citet{Ray2016}, who showed that language models
can be used to highlight buggy code. Indeed, we find that
our open-vocabulary NLM is more effective than previous LMs at highlighting buggy code.

More broadly, these contributions may impact
future development software tools.
First, source code LMs have been used in
a diverse variety of tools well beyond the obvious
application of autocompletion, ranging from code 
readability \cite{Allamanis2014} to program repair
\cite{chen2018sequencer}.
Our improved NLM could lead to improvements to all of these tools.
Second, recent results in NLP \cite{howard2018universal, Peters2018, Devlin2018} show that NLMs can be used as upstream tasks in transfer learning, leading to state-of-the-art improvement in downstream tasks: for instance, a model can be pre-trained as an NLM, and later on fine-tuned as a classifier.
Improved NLM architectures could lead to improved downstream classifiers, especially if the labelled data is scarce. While transfer learning from language models has been applied in software engineering \cite{robbes2019leveraging}, it has not been applied to source code due to the aforementioned vocabulary issues.
Finally, the general insights about vocabulary design that we study are not specific to NLMs, but arise whenever we build development tools by applying NLP methods to source code.

We conclude the paper in Section \ref{sec:conclusion}, and briefly describe the artifacts used in this work and how to obtain them in Section \ref{sec:artifacts}.

\section{Background and Related Work}
\label{sec:background}

We first note that this work is a consolidation of two unpublished works originally conducted independently: one work focused on the impact of various vocabulary choices on the resulting vocabulary size and the training of NLMs \cite{babii2019}, while the other work investigated the specific vocabulary choice of Byte-Pair Encoding, and introduced several improvements to the training procedure \cite{karampatsis2019}. This paper contains joint work that improves on both earlier works by investigating additional characteristics of the vocabulary, additional improvements to NLM training, an additional use case for NLMs, and a more thorough empirical evaluation. 

\subsection{Language Modeling in NLP}

A language model (LM) estimates the probabilities of sequences of words based on a training corpus. In NLP, these models have been applied to tasks such as speech recognition \cite{creutz2007morph} and machine translation \cite{jean2014using}.
Early language models were based on $n$-grams: the probability of a token is computed based on the $n-1$ previous tokens in the sequence.
These had success in NLP applications, but have two issues. First, they operate on small amounts of previous context, with $n$ often ranging from  3 to 6 (e.g. $n=6$ for
Java \cite{Hellendoorn2017}). Increasing $n$ does not scale well if the vocabulary is large: for a vocabulary of size $m$, there are $m^n$ possible n-grams. Second, they suffer from data sparsity: not all possible $n$-grams exist in the corpus. Smoothing \cite{chen1999empirical} alleviates---but does not eliminate---the issue.

 The current state-of-the-art in NLP is 
\emph{neural language models (NLM)} \cite{Bengio2003}. NLMs represent words in a continuous vector space, such that
words that are semantically similar are close in vector space \cite{Mikolov2013}, allowing the model to infer relationships between words, even if they do not appear in a specific context during training. This allows these models to better deal with data sparsity,
leading to better performance. 
Current NLMs are
based on architectures such as recurrent neural networks (RNN) \cite{Mikolov2010}, long short-term memory (LSTM) \cite{Hochreiter1997}, or Transformer \cite{vaswani2017attention} 
that model \emph{long range} dependencies: a study of LSTM NLMs showed that they use context as large as 250 words \cite{khandelwal2018sharp}, much longer than $n$-grams.


\subsection{Difficulties with Large Vocabularies}
ML models in general, and NLMs in particular, do not handle large vocabularies well. This is for several reasons:

\textit{Scalability.}
During pre-processing, each word is converted to a numerical representation, first via one-hot-encoding, producing (sparse) vectors of length equal to the vocabulary. NLMs then convert these to word embeddings, dense word vectors of much smaller dimensions (usually in the hundreds), in their first layer. For a vocabulary of size $m$ and embeddings of size $n$, the embedding layer is a dense matrix of size $m \times n$. A large $m$ (e.g., 100,000 or more) affects the memory required by the model
as well as the amount of computation required for training.
The output of an NLM is a prediction over the next token, 
which is a probability distribution over the entire vocabulary.
This must be computed once for each token in the training corpus many times during training.
This can be prohibitively slow for large vocabularies \cite{bradbury2016quasi,jozefowicz2016exploring}.


\textit{Out-of-vocabulary (OOV).} In traditional, \emph{closed-vocabulary} models, the vocabulary must be known in advance and will be built based on the training corpus. Any new word encountered at test time, called out-of-vocabulary
words, will not be able to be one-hot encoded as the resulting vector would exceed the expected dimensions. 
A common workaround is to have a specific \emph{unknown} token, and replace any word not previously seen by this token. This loses information, making the NLM unable to predict any new token, which is particularly problematic for source code.

\textit{Rare Words.}
Deriving meaningful embeddings for rare words is difficult since there is very little data to work with. Gong \etal show that the property that semantically similar words have similar embeddings does not hold for rare words: they hypothesize that since the words are rarely seen, the embeddings are rarely updated and thus stay close to their initialized values \cite{gong2018frage}. This issue is likely to impact performance: 
a very large vocabulary has been shown to negatively impact it, particularly with OOV words \cite{jean2014using}.


\subsection{Handling Large Vocabularies in NLP}


An \emph{open vocabulary model} is not restricted to a fixed-sized vocabulary determined at training time.
For instance, a character LM predicts each word letter by letter: its vocabulary is the set of characters; the OOV issue vanishes. However, it needs to model longer dependencies than a word NLM, impacting performance. Models using \emph{subword units}, or \emph{subwords},
combine the strengths of character and token LMs. A subword unit is a sequence of characters that occurs as a subsequence of some token in the training set; the model outputs a sequence of subword units instead of a sequence of tokens.
Many NLP models have used linguistically-motivated subwords \cite{creutz2007morph,Luong2013,Bazzi2002,Mikolov2012}.
Mikolov \etal found that subword models improved on character models \cite{Mikolov2012}. Sennrich \etal adapt the Byte-Pair Encoding (BPE) algorithm to decompose words in subwords, improving rare word translation \cite{Sennrich2015}. 
Kim \etal combine a character CNN with a NLM \cite{kim2016character}. Vania and Lopez compare LMs (words, morphs, character n-grams, BPE) on several languages \cite{Vania2017}. 



Another approach  to the OOV problem are \emph{cache models and copy mechanisms} \cite{merity2016pointer,Allamanis2016,grave2016improving}, which
allow the model to re-use words that have appeared previously. This helps with the OOV problem,
because such models can copy words that are not in their fixed vocabulary, but it does not help the \emph{first} time
an OOV word appears.

\subsection{Language Modeling and Vocabulary in SE}

\textit{Language Models in Software Engineering (SE).} Seminal studies have laid the groundwork for the use of language models on source code: Gabel and Su show that software is very repetitive \cite{Gabel2010}, which motivates
the use of statistical modelling for code.
\citet{Hindle2012} 
build language models of source code, finding applications in code completion. Nguyen et al. \cite{Nguyen2013} augmented $n$-gram LMs with semantic information such as the role of a token in the program, e.g., variable, operator, etc. Tu \etal \cite{Tu2014} find that software is even more repetitive taking local context into account. 
Rahman \etal find that while some aspects of software are not as repetitive as previously thought (non-syntax elements), others are even more so (API sequences) \cite{rahman2019revisiting}.
Other models of source code include probabilistic higher order grammars (PHOG) \cite{Bielik2016}, which use ASTs,
and several types of  RNNs, including LSTMs \cite{White2015, Dam2016, Hellendoorn2017}.

\textit{SE Applications of Language Models.}
Probabilistic code models have enabled many applications in software engineering (see \citet{big-code-survey} for a survey).
One example is recommender systems aiming to aid developers in writing or maintaining code: Hindle \etal used a token-level LM for code completion \cite{Hindle2012}, while later, Franks \etal improved on performance with Tu's cache \cite{Tu2014} and built a code suggestion tool for Eclipse  \cite{Franks15}.
Another application are recommendation systems for variable, method, and class names \cite{Allamanis2014, Allamanis2015a, Allamanis2016} that employ relevant code tokens as the LM context.
Campbell \etal \cite{Campbell2014} used $n$-gram language models to detect syntax error locations in Java code, and later used an NLM for the same purpose \cite{santos2018syntax}. 
Ray \etal \cite{Ray2016} showed that buggy code has on average lower probability than correct code, and that LMs can spot defects as effectively as popular tools such as FindBugs. 

Several approaches use neural machine translation, in which an encoder LM is paired to a decoder LM. Examples include recovering names from minified Javascript code \cite{vasilescu2017recovering, Bavishi2018}, or from decompiled C code \cite{jaffe2018meaningful}. Other applications include program repair \cite{chen2018sequencer}, learning code changes \cite{Tufano2019Learning}, or generating source code comments \cite{Hu2018}.
Gu \etal \cite{Gu2016} generate API usage sequences for a given natural language query.  They then learn joint semantic representations of bilingual API call sequences to support API call migration  \cite{gu2017deepam}. Yin \etal \cite{yin2018learning} mine pairs of natural language and code from Stack Overflow to support tasks such as code synthesis from natural language.

\textit{Large vocabularies in SE.}
The majority of models of source code used closed vocabulary models.
Hellendoorn and Devanbu rightly notice that NLMs trained on a software corpus would struggle due to vocabulary size \cite{Hellendoorn2017}, because identifiers, which are the bulk of source code, can be arbitrarily complex, and are often compound words (e.g., \texttt{thisIdentifierHas6WordsAnd2Numbers}), causing an explosion of possible identifiers.
To produce an NLM that can be trained in a reasonable amount of time, Hellendoorn and Devanbu impose drastic restrictions
which would be expected to reduce predictive accuracy,
restricting the training set to 1\% of the original corpus \cite{Allamanis2013} and the vocabulary to only include words which occur more than 5 times. 
Even so, the resulting vocabulary size is still exceeds 76,000 words.
Similarly, \citet{Pradel2018} had a large vocabulary of 2.4 million unique tokens: they limited it to the 10,000 most common tokens to reduce inaccuracies due to rare words. 

To limit this issue, previous work has segmented identifiers via a heuristic called
\emph{convention splitting}, which splits identifiers on camel case and underscores \cite{Allamanis2015a}. Even though this segmentation can handle \emph{some} OOV tokens, it is limited to combinations of subtokens appearing in the training set and thus unable to achieve a truly open vocabulary. Additionally, many of these subtokens are still infrequent, which hinders the model's ability to assign high scores to their compositions. For example, despite using convention splitting, the implementation of code2seq from Alon \etal \cite{alon2018code2seq} only keeps the 190,000 most common vocabulary words.

Several studies have empirically compared different techniques for automatically splitting identifiers \cite{enslen2009mining,hill2014empirical}.  These works consider the somewhat different problem of splitting identifiers into words in a way that matches human judgment. Subword units may not necessarily produce words that humans recognize, but they can be trivially reassembled into complete tokens before they are shown to a developer. Stemming  \cite{Willett2006} has also been used to reduce the number of vocabulary words by only keeping their roots; this is however destructive. Malik \etal combined convention splitting and stemming for type prediction \cite{malik2019nl2type}.

Few SE approaches use caches. Tu \etal \cite{Tu2014} and Hellendoorn and Devanbu \cite{Hellendoorn2017} use $n$-gram caches. Li \etal augment an RNN with a copy mechanism based on pointer networks \cite{Vinyals2015} to improve OOV code completion \cite{Li2018}. While it can reuse an OOV word after seeing it, it cannot predict the word's first use, learn its representation, or learn its dependencies, 
unlike our model. Copy mechanisms were also used for program repair \cite{chen2018sequencer}, and method naming \cite{Allamanis2016}.

\section{Datasets}
\label{sec:datasets}

We use code corpora from three popular programming languages: Java, C, and Python. 
We choose these languages because they have differences that could affect the performance of LMs. Java has extensively been used in related work \cite{Hindle2012,Allamanis2013,Nguyen2013,Tu2014,Dam2016,Hellendoorn2017}. Unlike Java, C is procedural, and makes it possible to write very terse code.\footnote{For examples, see \url{https://www.ioccc.org/}.} Python is a multi-paradigm dynamic language with little use of static typing. 
For Java we used the Java Github corpus of Allamanis \etal \cite{Allamanis2013}, also used in \cite{Hellendoorn2017}. The C and Python corpora were mined following the procedure described in \cite{Allamanis2013}; the C corpus was mined in \cite{Dudoladov2013} and the Python corpus was mined in \cite{Fiott2015}. 
For lexical analysis we used the Java lexer implemented in \cite{Hellendoorn2017}\footnote{https://github.com/SLP-team/SLP-Core}; for C and Python we used the Pygments\footnote{http://pygments.org/docs/lexers/} library. Descriptive statistics are in Table \ref{tab:datasets}.

For Python and C we sampled 1\% of the corpus for validation and 1\% for testing. Another 10\% of the corpus was sampled as a separate data set to learn subword encodings with BPE. The rest of the data was used for training. We also report results on a smaller subset of 2\% of our full training set. For Java, we used a slightly different procedure to make our experiment comparable to a previous study \cite{Hellendoorn2017}. We divide the data into five subsets as in the other two languages. The validation and test sets are the same as in \cite{Hellendoorn2017}, and our ``small train'' set is the same as their training set. To obtain the full Java train set, we collect all of the files in the Java Github corpus that do not occur in the validation or test set. Of these, we sampled 1000 random projects for the subword encoding data set, and the remaining projects were used as the full train set.

In the vocabulary study, both
training sets and test sets are used.
To train LMs, we preprocess the corpora to match \cite{Hellendoorn2017}, replacing non-ASCII character sequences such as Chinese ideograms inside strings with a special token (\nonEnglish), removing comments, and keeping strings. Note that the lexer in \cite{Hellendoorn2017} replaces all strings with length of 15 characters or more
with the empty string. In Python, we do not add any special tokens to represent whitespace. 

\begin{table}
\caption{Corpus statistics for each code corpus.}
\label{tab:datasets}
\resizebox{\linewidth}{!}{
\centering
\ra{1}
\begin{tabular}{@{}rrrcrrcrr@{}}
\toprule & \multicolumn{2}{c}{\textbf{Java}} & & \multicolumn{2}{c}{\textbf{C}} & & \multicolumn{2}{c}{\textbf{Python}} \\
\cmidrule{2-3} \cmidrule{5-6} \cmidrule{8-9} &   Tokens & Projects &&   Tokens & Projects &&   Tokens & Projects \\ \midrule
Full                                         &    1.44B &    13362 &&    1.68B &     4601 &&    1.05B & 27535 \\
Small                                        &   15.74M &      107 &&   37.64M &     177  &&   20.55M &   307 \\
BPE                                          &   64.84M &     1000 &&  241.38M &     741  &&  124.32M &  2867 \\
Valid.                                       &    3.83M &       36 &&   21.97M &     141  &&   14.65M &   520 \\
Test                                         &    5.33M &      38  &&   20.88M &      73  &&   14.42M &   190 \\
\hline 
\end{tabular}
}
\end{table}

\input{vocabulary-study.tex}

\section{Neural Language Model for Code}
\label{sec:rnnlm}



%

We present our NLM for code based on subword units, which is based on a Recurrent Neural Network (RNN). RNN LMs scan an input sequence forward one token at a time, predicting a distribution over each token given all of the previous ones. RNNs with gated units can learn when to forget information from the hidden state and take newer, more important information into account \cite{Hochreiter1997}. Among various gated units, GRUs \cite{Cho2014b} have been shown to perform comparably to LSTMs \cite{Hochreiter1997} in different applications \cite{Chung2014}.

We intentionally selected a small model as our base model: a single layer GRU NLM built upon subword units learned from BPE (Section~\ref{sec:bpe:0}). For each vocabulary entry we learn a continuous representation of 512 features, while the GRU state is of the same size. In all our experiments we used a learning rate of 0.1, dropout of 0.5  \cite{Srivastava14} and a maximum of 50 epochs of stochastic gradient descent with a minibatch size of 32 (for the small training sets) or 64 (for the full training sets). These hyper-parameters were tuned on the small train and validation sets. After each iteration we measure cross entropy on a validation set (Section~\ref{sec:intrinsic}). If the cross entropy is larger than the previous epoch then we halve the learning rate and this can happen for a maximum of 4 times, otherwise training stops. During training of the global model we unroll the GRU for 200 timesteps, following \cite{khandelwal2018sharp}. Our implementation is open source 
(GitHub URL omitted for review). 
We also experiment with larger capacity models (2048 hidden features and GRU state). 

\subsection{Selecting Subword Units with BPE}
\label{sec:bpe:1}


In our code LM, we address vocabulary issues by having the model predict \emph{subwords} rather than full tokens at each time step. 
Subwords are inferred by  BPE (Section~\ref{sec:bpe:0}) on a held out dataset of projects that are separate from the training, validation, and test sets. We experimented with three encoding sizes, i.e., the maximum number of merge operations: 2000, 5000, and 10000.
To train the LM, we first segment the train, validation, and test sets using the learned encoding. We transform each token into a character sequence, adding \eot after every token. Then we apply in order the merge operations from BPE to merge the characters into subword units in the vocabulary.\footnote{We use the BPE implementation from \url{https://github.com/rsennrich/subword-nmt}} As in \cite{Sennrich2015} we do not merge pairs that cross token boundaries. 
Finally, we train and test the NLM as usual on the data segmented in subword units.

\subsection{Predicting Tokens from Subword Units}
\label{sec:beam-search}

Autocompletion algorithms present a ranked list of $k$ predicted tokens rather than a single best prediction. With a model based on subword units, it is not obvious how to generate the top $k$ predictions, because a single token could be made from many subword units. We approximate these using a custom variation of the beam search algorithm. 
If the beam is large enough the algorithm can give a good approximation of the top-$k$ complete tokens.

The NLM defines a probability $p(s_1 \ldots s_N)$ for any subword unit sequence. The goal of the beam search is: given a history $s_1 \ldots s_N$ of subword units that already appear in a source file, predict the next most likely \emph{complete token}. A \emph{complete token} is a sequence of subword units $w_1 \ldots w_M$ that comprise exactly one token: that is, $w_M$ ends with \eot and none of the earlier subword units do. Beam search finds the $k$ highest probability complete tokens, where we denote a single token as the sequence of units $w_1 \ldots w_M$, that maximize the model's probability $p(w_1 \ldots w_M | s_1 \ldots s_N)$. Importantly, the length $M$ of the new complete token is \emph{not} fixed in advance, but the goal is to search over complete tokens of different length.

Given a value of $k$ and a beam size $b$, the algorithm starts by querying the model to obtain its predictions of  possible subword units, ranked by their probability. 
The algorithm uses two priority queues: one called \id{candidates} which ranks the sequences of subword units that still need to be explored during the search,  and one called \id{bestTokens} which contains the $k$ highest probability complete tokens that have been expanded so far. Each candidate is a structure with two fields, \id{text} which is the concatenation of all the subword units in the candidate, and \id{prob} which is the product of the probabilities of each subword unit in the candidate. Both of the priority queues are sorted by the probability of the candidate.

In each iteration, the algorithm pops the $b$ best candidates from the \id{candidates} queue, expands them with one additional subword unit, and scores their expansions. 
If an expansion creates a token (the new subword unit ends with \eot) then it is pushed onto the token queue and the worst token is popped. This maintains the invariant that \id{bestTokens} has size $k$. If the new expansion is not a complete token, then it is pushed onto the \id{candidates} queue, where it can potentially be expanded in the next iteration.

\subsection{Caching}
\label{sec:caching}

We also implement a simple caching mechanism for our NLM to exploit the locality of source code, particularly previously defined identifiers. At test time, each time an identifier is encountered, the 5-token history that preceded it is added to a cache alongside it. Differently to n-grams, we do not store probabilities, as the NLM will compute them. If the current 5-token history exists in the cache, the identifiers that followed it are retrieved (this is in practice very small, usually 1 or 2 identifiers). These identifiers are then scored by the NLM, and their probabilities are normalized to 1. The beam search described earlier is then run, and the two probability distributions are merged, according to a cache weight parameter: $cache\_pred \times cache\_weight + beam\_pred \times (1 - cache\_weight)$. The top 10 of the merged predictions are then returned.

We set the cache weight to 0.3. Note that, like beam search, this is a test-time only addition that does not affect training.

\subsection{Dynamic adaptation to new projects}
\label{sec:adaptation}

A \emph{global LM}, trained in a cross-project setting, will perform better if it is adapted to a new project \cite{Hindle2012, Tu2014}. LMs with n-grams also employ caches for this. Simply training an NLM from scratch on a new project will not have enough data to be effective, while training a new model on both the original training set and the new project would be impractical and computationally expensive.



Instead, we use a simple method of dynamically adapting our global NLMs to a new project. Given a new project, we start with the global NLM and update the model parameters by taking a single gradient step on each encountered sequence in the project after testing on it. This series of updates is equivalent to a single training epoch on the new project. (In our evaluations in Section~\ref{sec:evaluation}, we will split up the project files in such a way that we are never training on our test set.) We unroll the GRU for 20 time steps instead of 200 as in our global models, in order to update the parameters more frequently. We apply only one update for two reasons. First, it is faster, allowing the model to quickly adapt to new identifiers in the project. Second, taking too many gradient steps over the new project could cause the NLM to give too much weight to the new project, losing information about the large training set.



\section{Evaluation}
\label{sec:evaluation}


\label{sec:intrinsic}

\textbf{Intrinsic Evaluation: Language Modeling.} A good language model assigns high probabilities to real sentences and low probabilities to wrong ones. For code, fragments that are more likely to occur in human-written code should be assigned higher probability.
Precise scoring of code fragments is essential for tasks such as translating a program from one programming language to another \cite{Nguyen2013b, Karaivanov2014}, code completion \cite{Raychev2014, Franks15}, and code synthesis from natural language and vice versa \cite{Allamanis2015b, Desai2016, Nguyen2016, Raghothaman2016}. 

As in previous work, our intrinsic metric is the standard cross entropy. Cross entropy defines a score over a sequence of tokens $t_1$, $t_2$, ..., $t_{|C|}$. For each token $t_i$, the probability $p(t_i|t_1,...,t_{i-1})$ of each token is estimated using the model under evaluation. Then the average per token entropy is
$H_p(C) = -\frac{1}{|C|}\sum_{i=1}^{|C|}\log p(t_i |  t_1,...,t_{i-1}).$
Cross entropy is the average number of bits required in every prediction; lower values are better. It not only takes into account the correctness of the predictions, but also rewards high confidence.

Our NLMs define a distribution over subwords, not tokens. To compute cross entropy for subword NLMs, we segment each token $t_i$ into subwords $t_i = w_{i1} \ldots w_{iM}$. Then we compute the product
$ p(t_i |  t_1,...,t_{i-1}) = \prod_{m=1}^M p( w_{im} |  t_1,...,t_{i-1},w_{i1} \ldots w_{i,m-1}),$
where the right hand side can be computed by the subword NLM. This probability allows us to compute the cross entropy $H_p(C)$.

\textbf{Extrinsic evaluation: Code Completion.} 
We report the performance of our LMs on code completion, which is the task of predicting each token in a test corpus given all of the previous tokens in the file. We measure performance with mean reciprocal rank (MRR), as is common in code completion evaluation \cite{Bruch2009, Raychev2014, Tu2014, Hellendoorn2017}. Each time the LM makes a prediction, we get a ranked list of $k=10$
predictions. For each one, the reciprocal rank is the multiplicative inverse of the rank of the first correct answer. MRR is the average of reciprocal ranks for a sample of queries $Q$:
\begin{equation}
\mathit{MRR} = \frac{1}{|Q|}\sum_{i=1}^{|Q|}\frac{1}{rank_{i}}.
\label{eq:MRR}
\end{equation}
A simplified description of MRR is that it averages top-$k$ predictive performance across various $k$. 
Note that a correct suggestion at rank 1 yields an MRR of 1; at rank 2, 0.5; at rank 10, 0.1.
Thus, a small difference in MRR could indicate a large change in the ranked list, especially
for higher MRR values.

\textbf{Code Completion Scenarios.} \label{sec:scenarios}
We use three scenarios from previous work \cite{Hellendoorn2017}: Each \emph{static}, \emph{dynamic}, and \emph{maintenance} settings simulates a different way of incorporating NLMs in an IDE. The task is always to predict test set tokens, but the training sets differ: 

\paragraph{Static tests.} The model is trained on a fixed training corpus, and later evaluated on a separate test dataset. This is a cross-project setting: train, validation, and tests sets all contain separate projects. This simulates a single global LM that is trained on a large corpus of projects and then deployed to clients without adaption.

\paragraph{Dynamic tests.} In addition to the training set, the model can update its parameters \emph{after} it has made predictions on files in the test set (it never trains on test data). 
Our NLMs are adapted using the procedure described in Section \ref{sec:adaptation}. After each project, we restore the model to the global LM learned from the train set only. This simulates a setting in which some files from the project of interest are available for dynamic adaptation.

\paragraph{Software maintenance tests.} This scenario is even closer to real world usage, simulating everyday development where programmers make small changes to existing code. The LMs are tested on one file at a time in the test set. For each test file $F$, the train set plus all other files in the test project except $F$ is used as training data. As this requires retraining the NLM once per file in the test set, this scenario was previously deemed infeasible for NLMs in \cite{Hellendoorn2017}.


\paragraph{Identifiers only.} Recent work observed that LMs for completion perform worse on identifiers than other tokens \cite{hellendoorn2019code}. Therefore, we  also report
model performance, i.e. entropy and MRR,
on identifier tokens only (excluding primitive types).
To clarify differences between methods, we also report \emph{recall at rank 1 (R@1)},
the percentage of all identifier usages which are correctly
predicted at rank 1, and similarly recall at rank 10 (R@10), the percentage when the correct identifier
appears anywhere in the model's top $10$ predictions.


\section{Research Questions}
\label{sec:RQs}


\emph{RQ1. How does the performance of subword unit NLMs compare to state-of-the-art LMs for code?}
We compare subword unit NLMs to standard $n$-gram LMs \cite{Hindle2012}, cache LMs \cite{Tu2014}, state-of-the-art $n$-gram LMs with nested caching \cite{Hellendoorn2017}, token-level NLMs \cite{White2015}, and heuristic splitting NLMs \cite{Allamanis2015a}.
We do not compare with PHOG \cite{Bielik2016} and pointer network RNNs \cite{Li2018}: both do not have a full implementation available.
We do not evaluate character-level NLMs as they have not shown benefits for NLP. 


\emph{RQ2. Can subword unit NLMs scale to large code corpora? Does the additional training data improve performance?} 
\label{sec:rq2Definition}
Training on a larger corpus may improve a model's performance, but adding more data tends to have diminishing returns. After some point, a model's performance saturates. 
We evaluate if NLMs can make better use of large corpora than $n$-gram models. 
Moreover, training on larger data uses introduces scaling issues. 
Thus, performance in terms of runtime cost, memory usage, and storage becomes important. 

\emph{RQ3. How does the performance of subword unit NLMs vary across programming languages?}
In principle the learning methods for NLMs are language agnostic; however, the majority of studies evaluate only on Java. We check if code LMs are equally effective on other programming languages: C's terseness, or Python's lack of type information could negatively impact an LM's performance. 

\emph{RQ4. Is the dynamic updating effective to adapt subword unit NLMs to new projects?}
New projects introduce many new identifiers that do not appear even in a large cross-project corpus. An $n$-gram LM can exploit the strong locality that characterises code through caching \cite{Hindle2012, Tu2014}. 
Thus we ask whether NLMs can also benefit from dynamic adaptation via the procedure presented in Section~\ref{sec:adaptation}.\footnote{A naive approach to the software maintenance scenario retrains the model from scratch for every test file, which was rightly deemed infeasible for NLMs by \cite{Hellendoorn2017}} 
We compare our dynamic adaption technique against two dynamic $n$-gram models:
cache LMs \cite{Tu2014} and nested cache LMs \cite{Hellendoorn2017}.

\emph{RQ5. Are NLMs useful beyond code completion?} NLMs in NLP have shown to be useful in a variety of tasks, including translation or summarization; they have been recently shown to be state of the art in transfer learning. While testing all of these scenarios vastly exceeds the scope of this paper, we test whether NLMs improve upon n-gram LMs in the task of detecting buggy code \cite{Ray2016}.

\section{Results}
\label{sec:results}
Table \ref{tab:allResults} presents the evaluation metrics of all scenarios; we refer to it continuously. We used the $n$-gram implementation\footnote{\url{https://github.com/SLP-team/SLP-Core}, version 0.1} used in \cite{Hellendoorn2017} with the same parameters (n = 6); all NLMs are ours. We compute MRR on the first million tokens of the test set, as in \cite{Hellendoorn2017}. 
\input{bigtable-r1-r10.tex}


\begin{table}[tb]
\caption{Effect of vocabulary size on Java performance
of our open-vocabulary models (Python and C are similar).}
\footnotesize{
\label{tab:vocab_size_results}
\ra{1.1}
\small
\begin{tabular}{@{}lr @{\hspace{1em}} rrrrrrr @{}}
\toprule 
 \multicolumn{2}{c}{\multirow{2}{*}{\textbf{Vocab Size}\hspace{1ex}}} &  \multicolumn{2}{c}{\textbf{Static}} & \multicolumn{2}{c}{\textbf{Dynamic}} & \multicolumn{2}{c}{\textbf{Maint.}} & \textbf{Bugs} \\
& &  \textbf{Ent} & \textbf{MRR} & \textbf{Ent} & \textbf{MRR} & \textbf{Ent} & \textbf{MRR}  & \textbf{\% Ent $\downarrow$} \\
\midrule

\multicolumn{3}{l}{\textbf{Small Train}}\\
\hspace{1em} &
  2\,000             & 4.90 & 62.87 & 2.33 & 75.66 & 1.46 & 77.48 & 3.07 \\
& 5\,000             & 4.78 & 63.80 & 2.27 & 77.14 & 1.51 & 78.49 & 3.38 \\
& 10\,000            & 4.77 & 63.75 & 2.54 & 77.02 & 1.60 & 78.69 & 3.26 \\
\multicolumn{3}{l}{\textbf{Large Train}} \\
& 2\,000             & 3.59 & 68.87 & 1.84 & 77.69 & 1.03 & 78.85 & 4.09  \\
& 5\,000             & 3.35 & 69.87 & 1.72 & 79.18 & 1.06 & 80.31 & 4.71  \\
& 10\,000            & 3.15 & 70.84 & 1.72 & 79.94 & 1.04 & 81.16 & 4.92 \\
\bottomrule
\end{tabular} }
\end{table}

\subsection{RQ1. Performance of Models}

Because the full data set is so large, we compare the different variants of $n$-gram models against each other on the small Java training set,
and then we compare the best $n$-gram LM against our BPE NLM on the large Java data set. In Table~\ref{tab:allResults},
we see that 
the nested cache model has the best performance of the $n$-gram models,
with a large improvement over the simpler models (for example, improving MRR from 58\% to 77\% on Java against the basic $n$-gram model). 
This is consistent with the results of \cite{Hellendoorn2017}. However, our BPE NLM outperforms it.
(Note that cache models can not be evaluated in the static scenario since the cache would adapt to the test set).
Moving to the large data set, we find that the BPE NLM still outperforms the nested cache model, even though the nested cache model
was specifically designed for code.
While previous work \cite{hellendoorn2019code} found that closed NLMs underperformed on identifiers, we find that our BPE NLMs do not.
In the dynamic scenario, 74\% of identifiers are predicted within the top 10 predictions, with up to nearly 56\% in first position.

\textit{Open vs closed vocabulary.} To specifically evaluate the effect of relaxing the closed vocabulary assumption, we compare our open vocabulary NLM to two closed vocabulary NLMs: one that uses full tokens (Closed NLM), and another that splits tokens according to conventions (Heuristic NLM). Those models have otherwise the same architecture as the open vocabulary. In both cases, we find that the open-vocabulary NLM significantly outperforms both closed vocabulary NLMs, and can be trained even in the maintenance setting, unlike the closed versions. Of note, our closed vocabulary NLM performs better than the one in \cite{hellendoorn2019code}, as it utilizes a fully connected hidden layer and dropout.
Finally, in Table~\ref{tab:vocab_size_results}
we report the performance of the open vocabulary NLMs with different vocabulary sizes, obtained after 2000, 5000, and 10000 BPE merge operations. We see that performance on the small training set is similar across vocabulary sizes: a large vocabulary is not required for good performance.
 
\textit{Caches, and larger capacity.} Both our cache and increasing model capacity (from 512 to 2048 features) are beneficial, particularly for the identifiers. The cache improves MRR by 3 to 4\%, with more improvements for low ranks, which is especially important for completion. On the small corpus, the large model improves MRR by nearly 3\%, a smaller improvement than adding the cache. Both improvements are complementary, increasing identifier MRR by close to 6\%.
\conclusion{Open vocabulary NLMs are effective models of source code, even on a small corpus, yielding state of the art performance.}

\subsection{RQ2. Large Corpora}
We contrast performance between small and large training sets.

\textit{Leveraging data.} When trained on larger corpora, the performance of $n$-gram models (including nested cache variants) gets saturated and they are unable to effectively leverage the extra information  \cite{Hellendoorn2017}. In contrast, our model can better leverage the increase in training data when trained on the full corpus. In the static scenario, our NLMs decrease entropy by about 1.5 bits, while MRR increases by about 6\%. More data helps our NLMs learn to synthesize identifiers from subwords better and with higher confidence.

The improvements are smaller but still exist when the NLMs use dynamic adaptation: for all encoding sizes the entropy improves by 0.5 bits and MRR by 2 to 3\%. 
In contrast, the nested cache $n$-gram model entropy improves by less than 0.1 bits and MRR by less than 0.4\%. From that we conclude that subword unit NLMs can utilize a large code corpus better than $n$-gram models. As shown in Table \ref{tab:vocab_size_results}, larger training corpora tend to favor NLMs with larger vocabularies, particularly in terms of MRR; larger models leverage the additional data even better. For all models, the improvements are more visible for identifiers: the large train alone contributes close to 7\% of MRR for identifiers, versus 3\% overall for the NLM.  Finally, 
larger NLMs (2048 features) are even better at leveraging the additional training data, due to their increased capacity. Similarly, the cache still improves performance further, even with the large training set; both improvements complement each other. 



\textit{Resource usage.} While the nested cache $n$-gram model is competitive with Java identifiers, this comes at a significant cost: resource usage. Disk usage for $n$-gram models range from 150 to 500 Mb in the small training set to \textbf{6 to 8.5GB} in the large training set. RAM usage is even more problematic, as it ranges from around 5GB in the small training set, up to \textbf{50 to 60GB} in the large training set. \emph{This makes the large $n$-gram models unusable in practice as they exceed the memory requirements of most machines}. 

In contrast, the NLMs do not vary significantly with training set size; their size is fixed. They range from 15MB (BPE 2K) to 45MB (BPE 10K) on disk (up to 240MB for the large capacity models). RAM usage for NLMs vary between 2 to 4GB when training (and can be reduced at the expense of speed by reducing batch size), and is considerably lower at inference time (for actual code completion), ranging from 250 to 400MB. 
\emph{Thus, if we compare practically applicable models, the small NLM outperforms 
the small nested cache $n$-gram model by up to 5.13\% in identifier MRR, and up to 5.75\% recall at 1; the large NLM does so by 8.68\% (MRR), and 10.81\% (recall at 1)}.

The open vocabulary makes training NLMs on large corpora scalable as vocabulary ceases to grow with corpus size; training time scales linearly with added data. Our largest NLM (BPE 10k, 2048 features), can process around 350 to 550 hundred thousand tokens per minute (roughly 100 to 300 projects per hour depending on project size) on a consumer-grade GPU. This makes our dynamic adaptation procedure, which trains one project for one epoch, clearly feasible. Training the initial model is still a large upfront cost, but it takes from a day (small NLM) up to two weeks (large NLM) on our largest dataset, and needs to be performed once. At inference time, predicting 10 tokens with beam search takes a fraction of a second, fast enough for actual use in an IDE, even without additional optimization. This is not true for the closed models.
\conclusion{Open-vocabulary NLMs can scale; furthermore, they leverage the increased training data effectively. Large $n$-gram models do \emph{not} scale in terms of resources.}

\vspace{-1em}

\subsection{RQ3. Multiple Languages}
We contrast Java performance with Python and C. We see interesting differences between Java, Python, and C. First, $n$-gram models perform considerably worse in Python, while NLMs do very well. We hypothesize that this is due to the smaller size of Python projects in our corpus, which reduces opportunity for caching (the average Python project is 2 to 3 times smaller than the average Java project). C projects, on the other hand, are competitive with Java projects, particularly with caching; they are on average 2 times larger. Interestingly, the nested and nested cache n-gram models perform comparatively worse in C than in Java: C projects tend to have a flatter structure, rendering the nesting assumption less effective in this case. Finally, the (not applicable in practice) large n-gram model outperforms our NLMs for C. We observed anectodal evidence that there is considerable duplication in the C corpus, which may affect this result \cite{Allamanis2018}. For NLMs, the performance is more even across the board, with overall slightly worse performance for C, and somewhat better performance for Python.
\conclusion{Our NLM performance results hold for Java, C, and Python.}

\subsection{RQ4. Dynamic Adaptation}

We evaluate the effectiveness of our proposed method for adaption of NLMs in the dynamic and maintenance scenarios.
This is crucial for practical usage of NLMs, because the dynamic and maintenance scenarios simulate
the setting where the developer is modifying a large, existing project. Using within-project data provides a large performance boost:
Even though within each scenario, our NLMs outperform $n$-grams, most $n$-gram models in the dynamic scenario outperform NLMs in the static scenario.
The improvement due to dynamic adaptation is greater than the improvement due to an NLM. 
Of note, the situation in the large training set is different: the static large NLM trained on the large training set \emph{outperforms the  cache $n$-gram LMs in the dynamic scenario, and is competitive with it in the maintenance scenario},
in other words, our large data set is so large that it \emph{almost} makes up for not having within-project data,
but within-project information is clearly still crucial.

Once we apply the dynamic adaptation method to the NLMs, the picture changes. With dynamic adaptation, our model achieves better cross-entropy than the current state-of-the-art \cite{Hellendoorn2017}, making it an effective technique to fine-tune an NLM on a specific project.
Using this method, it is even possible to evaluate NLMs on the maintenance scenario, which was previously deemed infeasible by \cite{Hellendoorn2017} since multiple models had to be created, each trained on the entirety of the test set minus one file. 
This is possible for us because the combination of a small vocabulary size and our finetuning method running for only one epoch
make this scenario much faster.

\textit{Open vs closed NLMs.} Interestingly, the difference in performance between the open and closed vocabulary NLMs is larger in the dynamic setting. 
We hypothesize that dynamic adaptation helps the open-vocabulary model to learn project-specific patterns about OOV words; this is not possible for a closed vocabulary NLM.



\conclusion{Dynamic adaptation for NLMs yields the state of the art; 
static NLMs are competitive with some dynamic $n$-gram models, which bodes well for transfer learning.
}

\subsection{RQ5. Bug Detection}
Previous work has observed that $n$-gram language models can detect defects as they are less ``natural'' than correct code \cite{Ray2016}. In short, defective lines of code have a higher cross-entropy than their correct counterparts. To assess whether our code NLM is applicable beyond code completion, we compare the ability of different language models to differentiate between the two on the well-known \texttt{Defects4j} dataset \cite{just2014defects4j}.
Defects4J contains 357 real-world defects from 5 systems. Both a buggy and a corrected version of the system are provided and the changed lines can be extracted. 
We compute the difference in entropy between the buggy and the fixed version for each of the diff patches provided.
The extracted code snippets usually contains a few unchanged surrounding lines that provide useful context for the LMs.
We expect a better LM to have a larger entropy difference between the defective and the corrected version. 

We compute these metrics only for LMs in a static setting for three reasons: 1) we simulated the setting in which a bug detector is trained on one set of projects and used on unseen ones, 2) it is not clear how caches would be used in this scenario (should the LM ``know'' which file a bug is in?), and 3) doing so could involve training two LMs for each defect, which is very expensive.

The results are shown in the Java "bugs" column in Tables~\ref{tab:allResults} and \ref{tab:vocab_size_results}. As we hypothesized, open vocabulary NLMs feature a larger entropy drop for clean files than n-gram LMs or closed NLMs. The drop in entropy is 70\% to 100\% for the small training set, depending on vocabulary size and model capacity (larger is better). Furthermore, these models benefit from a large training set, with a larger drop of 127 to 173\%. 
We hypothesize that beyond data sparsity for identifiers, the NLM's long range dependencies are especially useful in this task.

\conclusion{Open-vocabulary NLM are better bug detectors than $n$-gram LMs, particularly when trained on large corpora.}

\section{Conclusions}
\label{sec:conclusion}

Source code has a critical difference with natural language: developers can arbitrarily create new words, greatly increasing vocabulary. This is a great obstacle for \emph{closed-vocabulary} NLMs, which do not scale to large source code corpora. We first extensively studied vocabulary modelling choices, and showed that the only viable option is an \emph{open-vocabulary} NLM; all other vocabulary choices result in large vocabularies, high OOV rates, and rare words.

We then presented a new open-vocabulary NLM for source code. By defining the model on subword units, which are character subsequences of tokens, the model is able to handle identifiers unseen in training while shrinking vocabulary by \emph{three orders of magnitude}. As a consequence, our NLM can scale to very large corpora: we trained it on data sets over a \emph{hundred times larger} than had been used for previous code NLMs. Our NLM also uses \emph{beam search}, \emph{dynamic adaptation}, and \emph{caching} to efficiently generate tokens and adapt to new projects. 
Finally, we showed that our NLM outperforms recent state-of-the-art models based on adding nested caches to $n$-gram language models for code completion and bug detection tasks, in a variety of scenarios, and in three programming languages.

Of course, this study has limitations: While we tried to be exhaustive and evaluated a large number of scenarios, we  could not evaluate all the possible combinations (hundreds) due to the resources needed, such as some large models or some large training scenarios. For this reason, we also refrained to evaluate other NLM architectures such as LSTMs \cite{Hochreiter1997}, QRNNs \cite{bradbury2016quasi}, Transformers \cite{vaswani2017attention}, or additional neural cache variants \cite{merity2016pointer, Vinyals2015}. For the same reason, as in \cite{Hellendoorn2017} we also limited MRR to 1 million tokens, which may cause discrepancies with entropy metrics as they are not evaluated on the same test set. We also limited ourselves to three languages, and did not fully evaluate the impact of code duplication \cite{Allamanis2018}.

We also hope that the simplicity and scalability will enable large capacity models for code, and the transfer learning opportunities they bring \cite{Devlin2018,radford2019language}; this has been explored in software engineering, albeit not for  source code \cite{robbes2019leveraging}.
Improved language models for code have the potential to enable new tools for aiding code readability \cite{Allamanis2014}, program repair  \cite{Ray2016,Campbell2014,Gupta17,Bhatia2016}, program synthesis \cite{gulwani2017program} and translation between programming languages \cite{Karaivanov2014,Nguyen2013b}.
Finally, the technique of using subword units is not limited to language modeling, but can easily be incorporated into any neural model of code, such as models to suggest readable names \cite{Allamanis2015a}, summarizing source code \cite{Iyer2016, Allamanis2016}, predicting bugs \cite{Pradel2018}, detecting code clones \cite{White2016}, comment generation \cite{Hu2018}, and variable de-obfuscation \cite{Bavishi2018}.

\section{Artifacts}
\label{sec:artifacts}
Several artifacts were used to conduct this study: data, source code, and models. To improve replication of this work, the specific version of each artifact used in this study can be referenced via a DOI. Table \ref{tab:dois} lists the DOI of each artifact. This paper can be referenced when any of these artifacts is used.

\textit{Datasets.} The datasets described in \ref{sec:datasets} were published in previous work: The Java corpus was produced by Allamanis \etal \cite{Allamanis2013}, and also used in \cite{Hellendoorn2017}. The C corpus was mined in \cite{Dudoladov2013} and the Python corpus was mined in \cite{Fiott2015}. We use the raw datasets for the vocabulary study, but preprocess them for NLM training. Further, we defined training and test sets for the C and Python corpora, and defined the large training set for the Java corpus.  

\textit{Source code.} We implemented the \emph{codeprep} library that supports a variety of pre-processing options for source code. We used \emph{codeprep} to gather the vocabulary statistics presented in section \ref{sec:choices}. Researchers that wish to use the library to pre-process source code for their own study can find the library at: \\ \url{https://github.com/giganticode/codeprep}. 

The open vocabulary language model described in \ref{sec:rnnlm}, as well as the scripts implementing the training procedure and the evaluation scenarios are available in the \emph{OpenVocabCodeNLM} library. Researchers wishing to extend our model can find it on GitHub at: \url{https://github.com/mast-group/OpenVocabCodeNLM}. 

\textit{Models.} The models that were trained and evaluated in section \ref{sec:results} are also made available for further use. Each model was trained on GPUs for periods ranging from a few hours, up to two weeks. These models can be used as-is for inference in a code completion scenario. Alternatively, they may be fine-tuned for other tasks, such as classification \cite{howard2018universal, robbes2019leveraging}.

\begin{table}[tb]
\caption{DOIs of artifacts used or produced by this work}
\label{tab:dois}
\small {
\centering
\begin{tabular}{ll}
\toprule
\textbf{Artifact} & \textbf{DOI} \\
\midrule
Java corpus & \url{\doijava} \\
C corpus & \url{\doic} \\
Python corpus & \url{\doipython} \\
\midrule
Java, pre-processed & \url{\doijavaprep} \\
C, pre-processed & \url{\doicprep} \\
Python, pre-processed & \url{\doipythonprep} \\
\midrule
codeprep &  \url{\doicodeprep} \\
OpenVocabCodeNLM & \url{\doinlm} \\
\midrule
Trained models & \url{\doimodels} \\
\bottomrule
\end{tabular}}
\end{table}

\section{Acknowledgements}

This work was supported in part by the EPSRC Centre for Doctoral Training in Data Science, funded by the UK Engineering and Physical Sciences Research Council (grant EP/L016427/1) and the University of Edinburgh. This work was partially funded by the IDEALS and ADVERB projects, funded by the Free University of Bozen-Bolzano. Parts of the results of this work were computed on the Vienna Scientific Cluster (VSC). 

\bibliographystyle{ACM-Reference-Format}
\bibliography{BPEPaperBib}

\end{document}

%% file: dois.tex
\def\doijava{https://doi.org/10.7488/ds/1690}
\def\doic{https://doi.org/10.5281/zenodo.3628775}
\def\doipython{https://doi.org/10.5281/zenodo.3628784}
\def\doicodeprep{https://doi.org/10.5281/zenodo.3627130}
\def\doinlm{https://doi.org/10.5281/zenodo.3629271}
\def\doimodels{https://doi.org/10.5281/zenodo.3628628}

\def\doijavaprep{https://doi.org/10.5281/zenodo.3628665}
\def\doicprep{https://doi.org/10.5281/zenodo.3628638}
\def\doipythonprep{https://doi.org/10.5281/zenodo.3628636}

%% file: vocabulary-study.tex
\section{Modeling Vocabulary}
\label{sec:choices}

We study a series of modeling choices for source code vocabulary. These choices may be implicitly made by researchers, with or without evaluating alternatives; they may not always be documented in their studies. By making the choices explicit, we can study their impact on the vocabulary.
We report results on Java; similar results can be observed for C and Python. Our evaluation criteria are:

\input{vocabulary-data.tex}

\begin{description}[leftmargin=*]
    \item[Scalability] 
Training of models should scale to thousands of projects.
Scalability is influenced by the vocabulary size (number of unique words) and the corpus size (number of tokens). We report both metrics on our full java training set, and compare them to a baseline with percentages. For instance: \vocab{11.6M, 100} and \corpus{2.43B, 100}.   
    \item[Information loss] Models should be able to represent the original input as much as possible; out-of-vocabulary (OOV) tokens are particularly undesirable. We build a vocabulary on the training set, and compare it with the test set vocabulary; we report the percentage of new vocabulary words seen in the test set. As large vocabularies do not scale, we report OOV for the unfiltered vocabulary, and a \underline{smaller vocabulary} (the 75,000 most frequent words, as in \cite{Hellendoorn2017}). To show trends, we also plot OOV for: full vocabulary, 200K, 100K, 75K, 50K, and 25K. Such as: \OOV{42}{79}{full}. 
    \item[Word frequency] As rare words have worse representations than frequent ones \cite{gong2018frage}, increasing word frequency is desirable.
Different modelling choices can increase or decrease the number of rare words.
We report the percentage of the vocabulary that has a frequency of 10 or less, and plot a bar chart showing the percentage of vocabulary with frequencies of 1000+, 1000--101, 100--11, 10--2, and 1. For instance: \freq{83}{full}.
\end{description}
\paragraph{Baseline:} \fullstats Our baseline is a vocabulary of unsplit tokens, except strings and comments that are split by whitespace (not doing so roughly doubles the vocabulary). This vocabulary is extremely large: more than 11 million unique words on Java-large. 
The OOV rate on the test set exceeds 40\% with the full vocabulary, showing that developers do create many new identifiers. The most common way to shrink vocabulary is to replace infrequent tokens with \unk. Doing so further worsens OOV issues: after reducing the vocabulary to a more manageable 75K, close to 80\% of the test vocabulary is unseen in the training set. Many words are infrequent: 83\% of vocabulary words have a frequency of 10 or less, with 25\% occurring only once.




\subsection{Filtering the vocabulary}
Simplest is to filter vocabulary items that are deemed less important. Filtering is destructive: it thus needs to be thoroughly justified.


\paragraph{English.} \asciistats Source code can contain many non-English words in identifiers, strings, and comments, either because developers use other languages, or for testing or internationalization purposes. 
Handling multilingual corpora is an NLP research topic in itself; we evaluate the simplifying assumption to limit a corpus to English. This is not trivial: dictionary-based heuristics have too many false positives (e.g. acronyms). We use a simple heuristic: a word is non-English if it contains non-ASCII characters. This is imperfect; ``caf\'e'', ``na\"ive'', or ``Heuristiken'' are misclassified. Non-English words are replaced with a \nonEnglish placeholder. Even then, the vocabulary shrinks by only 2\%, while OOV drops by only 3\% at 75K. 

\paragraph{Whitespace.} \nospacestats Some applications (e.g., pretty-printers [3]) may care about the layout of source code. Others may not, giving importance only to syntactic or semantic aspects (unless code layout is syntactically important, such as in Python). Filtering out whitespace reduces the vocabulary only by a handful of tokens, but reduces corpus size by 22\% (1.89B tokens).

\paragraph{Comments} \nocommentstats Comments often contain natural language, which is much less repetitive than code. While tasks such as detecting self-admitted technical debt \cite{Maldonado2017Using} rely on comments, others do not. Replacing comments by placeholder tokens (e.g., \tokcomment)
significantly reduces corpus size (a further 26\%), but its effect on vocabulary is limited (6\%, given that comments are already split on whitespace).

\paragraph{Strings.} \nostringstats
Similarly, string literals can be filtered, replacing them by a placeholder
token like \tokstring. This does not reduce corpus size as much (a further 5\%), but shrinks vocabulary a further 11\%, close to 9.5 million words. This is still extremely large. We also evaluate the configuration used in \cite{Hellendoorn2017}: strings are kept, unsplit, but strings longer than 15 characters are replaced by the empty string. For consistency with previous work, we use it as \textbf{new baseline}. It
increases vocabulary, OOV and infrequent tokens rate:
\HDstringstats  


\conclusion{Full token vocabularies range in the millions, and hence do not scale. OOV and frequency issues are extremely important.}

\subsection{Word Splitting}
Identifiers are the bulk of source code and its vocabulary. 
While new identifiers can be created at will, developers tend to follow \emph{conventions}. When an identifier is made of several words, in most
 conventions, the words are visually separated to ease reading, either in \texttt{camelCase} or in \texttt{snake\_case} \cite{binkley2009camelcase}. Thus, an effective way to reduce vocabulary is to \emph{split} compound words according to these word delimiters, as was done by Allamanis \etal \cite{Allamanis2015a}.

The decision whether to split compound words or not has important ramifications. First, it introduces additional complexity: the LM can no longer rely on the assumption that source code is a sequence of tokens. Instead,
compound words are predicted as a sequence of subtokens, albeit in a smaller vocabulary. 
Second, subtokens increase the length of the sequences, making it harder to relate the current subtokens to the past context, as it increases in size. This makes the approach unviable for $n$-grams as $n$ would need to increase significantly to compensate.

Splitting tokens has advantages: most obviously, the vocabulary can be much smaller. Consequently, the OOV rate is reduced. Third, a model may infer relationships between subtokens, even if the composed word is rare, as the subtokens are more common than the composed word. Finally, using subtokens allows a model to suggest \emph{neologisms}, tokens unseen in the training data \cite{Allamanis2015a}. 

\textit{Splitting.} \convsplitstats
Word splitting via conventions drastically reduces the vocabulary, by a close to an order of magnitude (slightly more than a million words), at the cost of increasing corpus size by 57\%. The impact on the OOV rate is also very large, as it decreases by a factor of 5 (in the unfiltered case; for a vocabulary of 75K it is a factor of 4). However, the effect on word frequency is limited, with only 3\% more words that are more frequent than 10 occurrences. 


\textit{Case.} \casesplitstats
Most commonly,
words in different case (e.g. \texttt{value}, \texttt{Value}, \texttt{VALUE}) will be distinct words for the LM.
This could increase the vocabulary, but removing case loses information.
A possible solution is to encode case information in separator tokens (e.g., \texttt{Value} becomes \Uppercase \texttt{value}; \texttt{VALUE} becomes \UPPERCASE \texttt{value}).
at the cost of increasing sequence length.
Case-insensitivity does decrease the vocabulary, but not by much (a further 2\%), while corpus size increases significantly (a further 30\%). Thus, our following configurations do not adopt it: our \textbf{new baseline}  keeps case.

\conclusion{Word splitting is effective, but the vocabulary is still large (a million words). OOV and frequency issues are still important.}


\subsection{Subword splitting}
As splitting on conventions is not enough, we explore further. 

\textit{Numbers.} \splitnumstats
Numeric literals are responsible for a large proportion of the vocabulary, yet \emph{their} vocabulary is very limited.  
Thus, an alternative to filtering them out is to model them as a sequence of digits and characters. 
This yields a considerable decrease in vocabulary with our previous baseline (37\%), for only 2\% increase in corpus size. For OOV, there is a slight improvement for a 75K vocabulary (2\%), as well as for frequency (28\% of words occur 10 times or more). 

\textit{Spiral.} \roninstats Several approaches exist that split a token into subtokens, but go beyond conventions by using Mining Software Repositories techniques, such as Samurai \cite{enslen2009mining}, LINSEN \cite{corazza2012linsen}, Spiral \cite{hucka2018spiral}, or even neural approaches \cite{markovtsev2018splitting}. We applied the Spiral token splitter, which is the state of the art. We observed a further 26\% reduction of the vocabulary, for a 2\% increase in corpus size compared to number splitting. Spiral was also very effective in terms of OOV, with 9\% of unseen word when the vocabulary is limited to 75K, and 3\% when unfiltered (476K words). The impact on frequency was limited. Even if this is encouraging, the OOV rate is still high. 

\textit{Other approaches.} Stemming \cite{Willett2006} can reduce vocabulary size,
but loses 
information: it is not always obvious how to recover the original word from its stem. We found that applying stemming can further reduce vocabulary by 5\%, which does not appear to be a worthwhile tradeoff given the loss of information.
Another option is
\textit{character models} that achieve an open vocabulary by predicting the source file one character a time. 
OOV issues vanish, but unfortunately, this drastically inflates sequence lengths, so a character model is not desirable. 

\conclusion{While these strategies are effective, they do not go far enough; vocabulary stays in the hundreds of thousands range. There are still OOV issues for unseen data; most words are uncommon.}

\subsection{Subword splitting with BPE}
\label{sec:bpe:0}

The final alternative we evaluate is subword segmentation via Byte-Pair Encoding (BPE). BPE is an algorithm originally designed for data compression, in which bytes that are not used in the data replace the most frequently occurring byte pairs or sequences \cite{Gage1994}. In subword segmentation, this corpus is represented as 
a sequence of subwords. Special end-of-token \eot symbols are added to allow us to convert from a sequence of subword units back into a sequence of tokens with ease. The approach was adapted to build NMT vocabularies \cite{Sennrich2015}: the most frequently occurring sequences of characters are merged to form new vocabulary words. 

BPE builds up the vocabulary of subwords iteratively,
at each iteration a training corpus is segmented according to the current vocabulary.
The initial vocabulary contains all characters in the data set and \eot, and the corpus is split into characters and \eot.  Then,
all symbol pairs appearing in the vocabulary are counted.  All the appearances of the most frequent pair $(S_1, S_2)$ are replaced with a unique new single symbol $S_1S_2$, which is added to the vocabulary, without removing
$S_1$ or $S_2$ (which may still appear alone). This procedure is called a merge operation.  The algorithm stops after a given maximum number $n$ of merge operations; this
is the only parameter.
The final output of the algorithm is (1) the new vocabulary, which contains all the initial characters plus the symbols created from the merge operations, and (2) the ordered list of merge operations performed in each iteration.
New data is segmented by splitting it into characters
and merging in the same order.

\begin{figure}
\footnotesize
{ \center Java Code:} \vspace{0em}\\
\begin{lstlisting}[basicstyle=\footnotesize]
public AttributeContext(Method setter, Object value) {
   this.value = value;
   this.setter = setter;
}
\end{lstlisting}
{\center Subword Units:} \vspace{0.5em}\\
\tt\raggedright\scriptsize{
public\eot\ Attribute\ Con\ text\eot\ (\eot\ Method\eot\ set\ ter\eot\ ,\eot\ Object\eot\ value\eot\ )\eot\ \{\eot\ this\eot\ .\eot\ value\eot\ =\eot\ value\eot\ ;\eot\ this\eot\ .\eot\ set\ ter\eot\ =\eot\ set\ ter\eot\ ;\eot\ \}\eot}

\caption{Example of Java code as a list of subword units.}
\label{fig:JavaCodeBPETokens}
\end{figure}


BPE has several advantages. First,  no word is OOV; the vocabulary
always contains all single characters, so unknown words at test time can be  represented using those subwords, if no longer subwords apply.
Second, the vocabulary dynamically adapts to the frequency of the sequences: 
common sequences will be represented by a single word (eg, \texttt{exception}), while rare ones will be segmented into more common subword units (such as roots, prefixes and suffixes); this helps with sparsity issues. Finally, BPE allows for fine-grained control of vocabulary size, by tuning the number of merge operations. A larger vocabulary will have more complete words and less sequences, smaller ones will have longer sequences. An example of a Java code snippet segmented into subwords is shown in Figure~\ref{fig:JavaCodeBPETokens}. We computed BPE for 1K, 2K, 5K, 10K and 20K merges, on a held-out set of 1K project. 


\textit{BPE Subwords.} \bpestats 
We apply BPE (10K merges) to our Java corpus with preprocessed as in \cite{Hellendoorn2017}, which we use as a baseline for comparison. 
As expected, the OOV issues vanish, even for an \emph{extremely small vocabulary}. The corpus size grows, but not more than previous choices we explored. Since BPE merges based on frequency, the resulting subtokens, no matter their size, are frequent: more than 97\% of the remaining words occur more than 1,000 times in the corpus, with very few words that are in the hundreds, and 1\% less than ten. Lower amounts of merges result in a smaller vocabulary, at the cost of a larger corpus size. Our largest BPE vocabulary, 20K, is 575 times smaller than our initial baseline; our smallest is 11,500 times smaller.\footnote{Note that including non-ASCII characters grows the vocabulary by $\approx$ 5,000 words in each case; a solution is to apply BPE at the \emph{byte} level, as done in \cite{radford2018improving}}



\textit{Qualitative examination.} While the goal of BPE is not to produce human-readable tokens, we examine how closely the splits BPE produces match human ones. We inspected 110 random identifiers, and provide anecdotal evidence of the types of splits produced by BPE. Our goal is not to provide strong evidence, but rather to give a sense to the reader of what BPE splits look like in practice. 

While some subwords are readable at BPE 1K 
(\texttt{File \spl Output \spl Service</t>}), some subwords are not  
(\texttt{Default \spl M \spl ut \spl able \spl Tre \spl e \spl Node</t>}), but look good at 5K 
(\texttt{Default \spl Mutable \spl TreeNode</t>}). BPE handles rare words gracefully, producing longer sequences of shorter units as expected. Some examples include rare words due to typos  
(\texttt{in \spl cul \spl ded \spl template</t>}) or foreign words 
(\texttt{v \spl orm \spl er \spl k \spl medi \spl en \spl au \spl f \spl lis \spl ter</t>}). Some rare words are split in root and suffix 
(\texttt{Grid \spl ify</t>}), but some acronyms may be unexpectedly split 
(\texttt{IB \spl AN</t>}). Further, BPE can split words correctly without case information 
(\texttt{http \spl client \spl lib</t>}, at 5K). 

\conclusion{BPE shrinks source code vocabulary \emph{very} effectively. Moreover, most of the vocabulary is frequent, improving embeddings.}

%% file: vocabulary-data.tex
\def\oovcolor{red}
\def\oovGraph#1{
  \begin{tikzpicture}[xscale=0.08, yscale=0.08]
    \useasboundingbox (-0.2,0.6) rectangle (6.2,4.2); 
    \begin{scope}[ycomb, yscale=0.3]
      \draw[fill=\oovcolor!100!black!10, \oovcolor!100!black!10] (0,0) rectangle (6,10); 
      \draw[xshift=-14pt,  \oovcolor!70!black!80, line width=1.15] plot[] file {#1}; 
      \draw[\oovcolor!100!black!25, line width=0.2] (-0.1,-0.4) rectangle (6.1,10.4);  
   \end{scope}
\end{tikzpicture}
}

\def\freqcolor{blue}
\def\frequencyGraph#1{
  \begin{tikzpicture}[xscale=0.08, yscale=0.08]
    \useasboundingbox (-0.2,0.6) rectangle (5.2,4.2); 
    \begin{scope}[ycomb, yscale=0.3]
      \draw[fill=\freqcolor!100!black!10, \freqcolor!100!black!10] (0,0) rectangle (5,10); 
      \draw[xshift=-14pt,  \freqcolor!70!black!80, line width=1.15] plot[] file {#1}; 
      \draw[\freqcolor!100!black!25, line width=0.2] (-0.1,-0.4) rectangle (5.1,10.4);  
   \end{scope}
\end{tikzpicture}
}

\def\vocab#1{#1\%}
\def\corpus#1{#1\%}
\def\OOV#1#2#3{#1\%, \underline{#2\%}, \oovGraph{oov-#3}}
\def\freq#1#2{#1\%, \frequencyGraph{freq-#2}}

\def\vocabstats#1#2#3#4#5#6{
\vocab{#2} \spl \corpus{#3} \spl \OOV{#4}{#5}{#1} \spl \freq{#6}{#1} \\ \noindent
}

\def\fullstats{
\vocabstats{full}{11.6M, 100}{2.43B, 100}{42}{79}{83}
}

\def\asciistats{
\vocabstats{ascii}{11.4M, 98}{2.43B, 100}{35}{76}{83}
}

\def\nospacestats{
\vocabstats{ascii}{11.4M, 98}{1.89B, 78}{35}{76}{83}
}

\def\nocommentstats{
\vocabstats{nocomments}{10.8M, 93}{1.26B, 52}{38}{78}{83}
}

\def\nostringstats{
\vocabstats{nostrings}{9.5M, 82}{1.15B, 47}{39}{78}{83}
}

\def\HDstringstats{
\vocabstats{HDstrings}{10.9M, 94}{1.15B, 47}{39}{80}{84}
}

\def\convsplitstats{
\vocabstats{convsplit}{1.27M, 12}{1.81B, 157}{8}{20}{81}
}

\def\casesplitstats{
\vocabstats{casesplit}{1.09M, 10}{2.16B, 187}{9}{21}{83}
}

\def\splitnumstats{
\vocabstats{splitnum}{795K, 63}{1.85B, 102}{6}{18}{72}
}

\def\roninstats{
\vocabstats{ronin}{476K, 37}{1.89B, 104}{3}{9}{70}
}

\def\stemstats{
\vocabstats{stem}{410K, 32}{1.89B, 104}{3}{8}{74}
}

\def\bpestats{
\vocabstats{bpe}{10K, 1}{1.57B, 137}{0}{0}{1}
}

%% file: bigtable-r1-r10.tex
\newcommand{\one}[1]{\textbf{#1}}
\newcommand{\two}[1]{\underline{#1}}

\begin{table*}[tb]
\caption{Performance of the various models (bold: best, underlined: second best).}
\label{tab:allResults}
\centering
\ra{1.3}
\resizebox{\textwidth}{!}{
\begin{tabular}{@{}ll c@{\hspace{2pt}} rrrrrrr c@{\hspace{2pt}}  rrr  c@{\hspace{2pt}}  rrrr  c@{\hspace{2pt}}  rrrr@{}}
\toprule 
& \multirow{3}{*}{\textbf{MODEL}}  & \hspace{2pt}  & \multicolumn{7}{c}{\textbf{Java}} & \hspace{2pt} & \multicolumn{3}{c}{\textbf{Java Identifiers}} & \hspace{2pt} & \multicolumn{4}{c}{\textbf{C}} & \hspace{2pt} & \multicolumn{4}{c}{\textbf{Python}} \\
\cmidrule{4-10} \cmidrule{12-14} \cmidrule{16-19} \cmidrule{21-24}
& & & \multicolumn{2}{c}{\textbf{Static}} & \multicolumn{2}{c}{\textbf{Dynamic}} & \multicolumn{2}{c}{\textbf{Maintenance}} & \textbf{Bugs} & & 
      \multicolumn{3}{c}{\textbf{Dynamic}} & &
      \multicolumn{2}{c}{\textbf{Static}} & \multicolumn{2}{c}{\textbf{Dynamic}} & &
      \multicolumn{2}{c}{\textbf{Static}} & \multicolumn{2}{c}{\textbf{Dynamic}} \\
\cmidrule{4-5} \cmidrule{6-7} \cmidrule{8-9}  \cmidrule{10-10}  \cmidrule{12-12} \cmidrule{13-14} \cmidrule{16-17} \cmidrule{18-19}   \cmidrule{21-22} \cmidrule{23-24}
& & & {Ent} & {MRR} & {Ent} & {MRR} & {Ent} & {MRR} & {\% Ent $\downarrow$} & &
      {R@1} & {R@10}& {MRR} & &
      {Ent} & {MRR} & {Ent} & {MRR} & &
      {Ent} & {MRR} & {Ent} & {MRR} \\
\midrule

\multicolumn{2}{l}{\textbf{Small Train}}\\
 & 
\textbf{$n$-gram}                 & & 6.25 & 53.16 & 5.54 & 56.21 & 5.30 & 58.32 & 1.81 & & 17.24 & 34.66 & 22.26 & & 6.51 & 55.20 & 4.14 & 57.34 & & 5.30 & 43.63 & 4.81 & 47.39  \\
& \textbf{Nested}                 & &  -   &   -   & 3.65 & 66.66 & 2.94 & 71.43 &   -  & & 37.46 & 56.85 & 43.87 & &   -  &   -   & 3.61 & 62.25 & &   -  &   -   & 4.05 & 54.02 \\
& \textbf{Cache}                  & &  -   &   -   & 3.43 & 69.09 & 3.32 & 70.23 &   -  & & 40.13 & 59.52 & 46.57 & &   -  &   -   & 2.19 & 75.09 & &   -  &   -   & 3.22  & 62.27 \\
& \textbf{Nested Cache}           & &  -   &   -   & 2.57 & 74.55 & \two{2.23} & \two{77.04} &   -  & & 49.93 & \two{70.09} & \two{56.81} & &   -  &   -   & 2.01 & 76.77 & &   -  &   -   & 2.89 &  65.97   \\
& \textbf{Closed NLM}             & & \one{4.30} & 62.28 & 3.07 & 71.01 &  -   &   -   & 1.81 & & 30.96 & 49.93 & 37.20 & & 4.51 & 60.45 & 3.20 & 72.66 & & 3.96 & \one{81.73} & 3.34 & 84.02 \\
& \textbf{Heuristic NLM}          & & \two{4.46} & 53.95 & 3.34 & 64.05 &  -   &   -   & 1.04 & & 39.54 & 58.37 & 45.28 & & 4.82 & 52.30 & 3.67 & 61.43 & & 4.29 & 65.42 & 3.56 & 71.35  \\
 & \textbf{BPE NLM (512)}         & & 4.77 & \two{63.75} & \two{2.54} & 77.02 & \one{1.60} & \one{78.69} & \two{3.26} & & 45.49 & 67.37 & 52.66 & & \two{4.32} & \two{62.78} & \two{1.71} & \two{76.92} & & \two{3.91} & 81.66 & \two{2.72} & \two{86.28} \\
 & \textbf{BPE NLM (512) + cache} & &  -   &   -   &  -   & \two{77.42} &  -   &   -   &   -  & & \two{50.49} & 68.16 & 56.30 & &   -  &   -   &   -  &   -   & &  -   &   -   &   -  &   -   \\
 & \textbf{BPE NLM (2048)}        & & 4.77 & \one{64.27} & \one{2.08} & 77.30 &  -   &   -   & \one{3.60} & & 48.22 & 69.79 & 55.37 & & \one{4.22} & \one{64.50} & \one{1.59} & \one{78.27} & & \one{3.66} &   \two{81.71}   &   \one{2.69}  &   \one{86.67}   \\
 & \textbf{BPE NLM (2048) + cache}& &  -   &   -   &  -   & \one{78.29} &  -   &   -   &   -  & & \one{52.44} & \one{70.12} & \one{58.30} & &   -  &   -   &   -  &   -   & &  -   &   -   &   -  &   -   \\
\midrule
\multicolumn{2}{l}{\textbf{Large Train}} \\
& \textbf{Nested Cache}          & &  -   &   -   & 2.49 & 75.02 & \two{2.17} & \two{77.38} &   -  & & 52.20 & 72.37 & 59.09 & &   -  &   -   & 1.67 & \one{84.33}  & &   -  &   -   & \one{1.45} & 71.22  \\
& \textbf{BPE NLM (512)}         & & \two{3.15} & \two{70.84} & \two{1.72} & 79.94 & \one{1.04} & \one{81.16} & \two{4.92} & & 51.41 & \two{74.13} & 59.03 & & \two{3.11} & \two{70.94} & \two{1.56} & 77.59 & & \two{3.04} & \two{84.31} & 2.14 & \two{87.06} \\
& \textbf{BPE NLM (512) + cache} & &   -  &   -   &   -  & 80.29 &   -  &   -   &   -  & & 55.68 & \one{74.30} & 61.94 & &   -  &   -   &    -  &   -   & &   -  &   -   &   -  &   -   \\
& \textbf{BPE NLM (2048)}        & & \one{2.40} & \one{75.81} & \one{1.23}  & \two{82.41} & - & -  & \one{5.98} & & \two{57.54} & 72.18 & \two{62.91}  && \one{2.38} & \one{80.17} & \one{1.36} & \two{83.24} &  & \one{2.09} & \one{86.17} & \two{1.90} & \one{87.59} \\
& \textbf{BPE NLM (2048) + cache} & &   -  &   -   &   -  & \one{83.27} &   -  &   -   &   -  & & \one{60.74} & 73.76 & \one{65.49} & &   -  &   -   &    -  &   -   & &   -  &   -   &   -  &   -   \\
\bottomrule
\end{tabular}
} 
\end{table*}